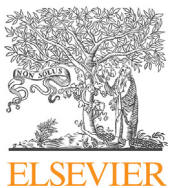
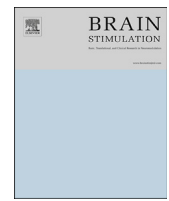

# Evidence of state-dependence in the effectiveness of responsive neurostimulation for seizure modulation

Sharon Chiang [a, *], Ankit N. Khambhati [b], Emily T. Wang [c], Marina Vannucci [c], Edward F. Chang [b], Vikram R. Rao [a]

[a] Department of Neurology and Weill Institute for Neurosciences, University of California, San Francisco, San Francisco, CA, United States
[b] Department of Neurological Surgery, University of California, San Francisco, San Francisco, CA, United States
[c] Department of Statistics, Rice University, Houston, TX, United States

## ARTICLE INFO

## ABSTRACT

*Background:* An implanted device for brain-responsive neurostimulation (RNS® System) is approved as an effective treatment to reduce seizures in adults with medically-refractory focal epilepsy. Clinical trials of the RNS System demonstrate population-level reduction in average seizure frequency, but therapeutic response is highly variable.

*Hypothesis:* Recent evidence links seizures to cyclical fluctuations in underlying risk. We tested the hypothesis that effectiveness of responsive neurostimulation varies based on current state within cyclical risk fluctuations.

*Methods:* We analyzed retrospective data from 25 adults with medically-refractory focal epilepsy implanted with the RNS System. Chronic electrocorticography was used to record electrographic seizures, and hidden Markov models decoded seizures into fluctuations in underlying risk. State-dependent associations of RNS System stimulation parameters with changes in risk were estimated.

*Results:* Higher charge density was associated with improved outcomes, both for remaining in a low seizure risk state and for transitioning from a high to a low seizure risk state. The effect of stimulation frequency depended on initial seizure risk state: when starting in a low risk state, higher stimulation frequencies were associated with remaining in a low risk state, but when starting in a high risk state, lower stimulation frequencies were associated with transition to a low risk state. Findings were consistent across bipolar and monopolar stimulation configurations.

*Conclusion:* The impact of RNS on seizure frequency exhibits state-dependence, such that stimulation parameters which are effective in one seizure risk state may not be effective in another. These findings represent conceptual advances in understanding the therapeutic mechanism of RNS, and directly inform current practices of RNS tuning and the development of next-generation neurostimulation systems.

## Introduction

Epilepsy is a common neurological disorder characterized by recurrent seizures. Anti-seizure medications are first-line treatments for epilepsy, but up to 30% of patients have seizures that are incompletely controlled with medications. For these patients with medically-refractory epilepsy, resection of the seizure focus can be highly effective but may not be possible if seizure foci are bilateral, poorly localized, spatially-extensive, or overlapping with eloquent cortex.

In recent years, responsive neurostimulation (RNS® System), brain electrical stimulation triggered by real-time detection of seizures, has emerged as a promising therapy for patients with medically-refractory epilepsy who are not candidates for resective surgery. The RNS System (NeuroPace, Inc.) is a cranially-implanted device approved as a safe and effective treatment for adults with medically-refractory seizures arising from one to two foci. In clinical trials, RNS System treatment results in median seizure frequency reduction of 75% after nine years [1,2]. However, therapeutic response to RNS is highly variable across individuals. Up to one-third of patients have a dramatic response to RNS with ≥90% seizure frequency reduction [2], but about one-quarter of patients are non-responders with less than 50% reduction in seizure frequency [1]. Investigations of factors that predict therapeutic response to RNS have focused primarily on stationary clinical features, such as the presence of mesial temporal sclerosis, seizure focus, prior resection, prior intracranial monitoring, or lead proximity to the seizure focus [3–5]. No features have yet been found to explain this variability [3–5]. Improving outcomes for all patients treated with RNS, particularly non-responders, may be facilitated by greater mechanistic understanding of this therapy and optimization of stimulation protocols based on individual temporal dynamics.

Prevailing views on the mechanistic basis of RNS have shifted over time. Originally, the efficacy of RNS was thought to derive from direct inhibition of ongoing seizures by electrical stimulation [6–8]; in other words, RNS was thought to abort seizures in the way that defibrillators terminate cardiac arrhythmias. However, a critical observation from long-term clinical trials is that the efficacy of RNS improves over time [1,2], suggesting chronic, neuromodulatory effects in addition to acute seizure-terminating effects. Still, many questions remain about how best to harness the putative neuromodulatory effects of RNS. These effects likely depend on the stimulation parameters that are employed [9,10], but aside from experience-based practice, there are virtually no objective methods for selecting these parameters rationally. For example, based on experience during clinical trials, charge density is often increased empirically to enhance therapeutic response, but a definitive correlation between charge density and seizure frequency has not been shown [5].

Emerging research indicates that epilepsy is a cyclical disorder. Rates of interictal epileptiform discharges (IEDs) fluctuate with daily and multi-day periodicities [11–18]. Electrographic seizures occur preferentially at certain phases of these IED cycles, which therefore help determine momentary seizure risk [18,19]. Periods of high and low seizure risk in IED fluctuations can be conceptualized as distinct brain states oscillating with hours-to-months long macro-periodicities, which may have varying susceptibility to neurostimulation. In vitro and in vivo models suggest that neural modulation using electrical stimulation is state-dependent [20,21]. Despite these discoveries, RNS System programming in contemporary practice remains empiric and agnostic to brain state. Standardized titration schedules for stimulation parameters are applied to patients with diverse epilepsies [22], and parameters are adjusted relatively infrequently. Thus, the dynamism of brain states in epilepsy, based on semi-chronic oscillations on the hourly or longer time scale of IEDs, contrasts sharply with clinical management of RNS. Whether electrical stimulation protocols can be tailored in a patient-specific, time-varying manner to increase response rates has not been established.

Here, we hypothesize that the ability of electrical stimulation to modulate seizures is state-dependent, and, specifically, that stimulation parameters that are effective in one brain state may not be effective in another brain state. To test this, we investigate the effect of electrical stimulation parameter changes on the incidence of electrographic seizures as a state-dependent function in a retrospective cohort of 25 adults with medically-refractory focal epilepsy treated with the RNS System.

## Material and methods

### Patient selection

We performed a retrospective review of patients with medically-refractory focal epilepsy who were implanted with the RNS System for clinical indications between August 2014 and June 2018 at the University of California, San Francisco (UCSF) Comprehensive Epilepsy Center. Data collection was approved by the UCSF Institutional Review Board and written informed consent was obtained from all subjects.

### Data collection

RNS System stimulation parameters (charge density, stimulation frequency, burst duration, pulse width, burst duration) and stimulation pathway (bipolar, monopolar, or grouped bipolar) were determined by treating providers based on clinical information. Examples of bipolar, grouped bipolar, and monopolar pathways are shown in Table 1. All patients had two four-contact depth and/or cortical strip leads placed in or over the seizure-onset zone(s). Computed tomography brain imaging was performed postoperatively to confirm lead placement. For each patient, the number of LE detections, stimulation settings (charge density, frequency, burst duration, pulse width), and stimulation pathway within each hour were recorded. Stimulation parameters used in the first therapy delivered by the device during each detection episode were analyzed.

RNS System stimulation settings and hourly LE counts were extracted from the Patient Data Management System (PDMS), a secure online repository of RNS System programming parameters and electrographic data. Stimulation pathways were analyzed separately. Less than 0.1% of the data were non-consecutive (i.e., break between recorded hourly counts). All related clinical data were abstracted from the electronic medical record system.

### State-dependent modeling of parameter associations with long episodes

Seizures are increasingly understood as stochastic realizations of fluctuating seizure risk, and computational models of seizure frequency and timing have sought to characterize underlying risk states [11,13–15]. We used a first-order Bayesian non-homogeneous hidden Markov model for zero-inflated count data to identify whether stimulation parameters differed in effectiveness depending on current risk state [11,12]. We adapted the method developed and validated by Refs. [11,12] to RNS System data, which proposed a Bayesian algorithm for a non-homogeneous hidden Markov model for identifying latent values of seizure risk and estimating state-dependent associations of external covariates with changes in ECoG [11,12]. Due to ECoG storage limitations on the neurostimulator, LE detections were used to identify electrographic seizures. All LE ECoG data were visually reviewed by an experienced epileptologist (V.R.R.) to determine how reliably they corresponded to electrographic seizures, as described previously [14,19,23,24]. Only patients for whom >90% of LE's detected by the RNS System corresponded to electrographic seizures [19] and for whom there were at least 20 time points (i.e. 1 hour time periods) in one of the stimulation configurations were included. The latent process model proposed by Refs. [11,12] was used to decode electrographic seizures into underlying latent (unobserved) levels of seizure risk, while simultaneously estimating the association of exogenous inputs at each time $t$ (here, RNS stimulation parameters) with changes in latent risk from time $t$ to $t+1$ (Supplementary Figure A1). For an in-depth description of the methods used here, we refer the reader to Refs. [11,25].

**Table 1**
**Responsive neurostimulation pathway configurations**. For each configuration, example stimulation pathways, accumulated person-hours, and sample size (N) are reported.

| | Definition | Example stimulation pathways[a] | Accumulated experience (person-hours) | Sample size (N) |
|---|---|---|---|---|
| Monopolar | One anodal electrode and one cathodal electrode | 1. (----)(++++)(0) <br> 2. (----)(0000)(+) <br> 3. (++++)(-----)0) <br> 4. (0---)(0000)(+) <br> 5. (00--)(++++)(0) <br> 6. (0000)(--)(+) | 150,238 person-hours | 13 |
| Bipolar | Alternating anodal and cathodal electrode contacts | 1. (+++-)(-+++)(0) <br> 2. (++-)(0000)(0) <br> 3. (0-+-)(0000)(0) <br> 4. (0++)(0000)(0) <br> 5. (0000)(+-+-)(0) <br> 6. (0000)(00+-)(0) | 130,266 person-hours | 16 |
| Grouped bipolar | Alternating pairs of anodal and cathodal electrode contacts | 1. (++--)(0000)(0) <br> 2. (0000)(--++)(0) <br> 3. (0000)(+++-)(0) | 5969 person-hours | 1 |

[a] Negative (-) denotes cathode; positive (+) denotes anode; first parentheses denote electrode contact designations for the first lead; second parentheses denote electrode contact designations for the second lead; third parentheses denotes neurostimulator canister.

Markov chain Monte Carlo (MCMC) sampling was used to sample from the posterior distribution of model parameters, which yields posterior estimates of the association of exogenous inputs with seizure risk state transitions. We analyzed the association of commonly adjusted RNS stimulation parameters—including charge density, pulse frequency, and burst duration—with state transitions. Pulse width was excluded from bipolar and monopolar analyses, as this parameter is less commonly manipulated than charge density, pulse frequency, burst duration, or stimulation pathway. Burst duration was included only in monopolar analysis, as the only value of burst duration used in patients employing a bipolar stimulation configuration was 100 ms. Burst duration analysis in this study therefore only pertains to monopolar stimulation. In addition, we controlled for several other baseline characteristics within the hidden Markov model, including age, sex, and seizure focus (mesiotemporal, neocortical, or both) [5]. To control for potential influence of device detection sensitivity on LE counts, the number of episode starts was additionally included as a covariate in the non-homogeneous hidden Markov model. The number of episode starts correlates with increased detection sensitivity and was used as a proxy for device programming-related changes in detection sensitivity, which allows for quantification of the marginal effect of electrographic seizure counts while considering the partial contribution of device detection sensitivity via episode starts. For each configuration, we explored model fits with $K$ between 2 and 3 states to find the number of states $K$ yielding the best model fit. Model fit for each value of $K$ was assessed using the deviance information criterion [26] and convergence of the state allocations to the stationary distribution. Models with lower DIC indicate better goodness of fit and are generally preferable to models with higher DIC. Convergence of each chain was assessed based on trace plots and the Raftery–Lewis diagnostic. Posterior means and 95% highest posterior density (HPD) intervals for multinomial logit coefficients were computed to estimate the association between neurostimulation parameter values with state transitions. Posterior estimates of states were computed based on the posterior mode. Full Bayesian significance testing was performed based on whether the 95% HPD interval contains zero [27]. Zero-valued coefficients mean that there is insufficient evidence to suggest that the stimulation parameter is associated with the improvement or worsening in seizure risk. Coefficients with non-zero values mean that there is evidence that the stimulation parameter is associated with the improvement or worsening in seizure risk. Subgroup analyses were additionally performed for responders and non-responders to RNS. Responders were defined as patients with 50% or greater reduction in seizure frequency, relative to pre-implant baselines [28].

## Results

From a total cohort of 49 patients implanted with the RNS System at our center between August 2014 and June 2018, we identified 25 patients with available data for whom LE reliably corresponded to electrographic seizures (see Methods). These patients represented a total accumulated experience of 286,473 patient-hours of treatment with the RNS System.

### Demographic characteristics

Demographic characteristics are provided in Table 2. The mean age at time of RNS System implantation was 38.7 ± 13.9 years (range, 18.2–72.0 years). A greater proportion of patients treated with bipolar stimulation had a mesiotemporal focus (p = 0.008), whereas a greater proportion of patients treated with monopolar stimulation had a neocortical focus (p = 0.009). Monopolar and bipolar cohorts were otherwise similar in age, sex, seizure frequency, age of epilepsy onset, epilepsy duration, epilepsy etiology, and 50% responder rates (Table 2). The average epilepsy duration was 21.6 ± 13.0 years (range 2.0–61.0 years). The most common etiology of epilepsy was cryptogenic (54.2%). The median charge density for neocortical leads across all person-hours was 1.3 μC/cm² (IQR, 0.2 μC/cm²). The median charge density for hippocampal leads across all person-hours was 2.0 μC/cm² (IQR, 1.3 μC/cm²). Stimulation frequency and burst duration were similar between mesiotemporal and neocortical leads (for mesiotemporal leads, median (IQR) frequency of 200 (50.05) Hz and burst duration 100 (1141.8) ms; for neocortical leads, median (IQR) frequency 200 (14.26) Hz and burst duration 100 (72.16) ms).

### RNS System parameters

RNS System stimulation electrodes were placed in the seizure onset zone(s) for all patients. For patients with regional neocortical seizure onset zones, strip leads were placed on the cortical surface at least 1 cm apart flanking the seizure onset zone while residing within it, as previously described [5]. For patients with hippocampal seizure onset zones, depth electrodes were placed via a trans-occipital trajectory with electrodes within the hippocampus along its long axis [29]. Stimulation was enabled on the RNS System once reliable seizure detection was achieved. Initial RNS System stimulation parameters were empirically selected and iteratively adjusted based on the patient's clinical course and experience from the RNS System Clinical Trials. A monopolar stimulation pathway was used in 13/25 patients and a bipolar stimulation pathway was used in 16/25 patients. The grouped bipolar stimulation pathway was used temporarily in only one patient, with the same stimulation settings used during the entire duration of grouped bipolar stimulation (charge density of 2.0 μC/cm², frequency of 200 Hz, pulse width of 160 μs, and burst duration of 100 ms) and was thus excluded from analysis. Of the 25 patients, 8 patients were treated only with monopolar stimulation, 12 patients were treated only with bipolar stimulation, four were initially treated with bipolar stimulation and then switched to monopolar stimulation, and one patient was treated initially with monopolar stimulation and then switched to grouped bipolar stimulation. A total of 150,238 person-hours from 13 patients were available for analysis in the monopolar stimulation configuration and 130,266 person-hours were available from 16 patients in the bipolar stimulation configuration.

**Table 2**
**Demographic characteristics of patients using each RNS stimulation configuration**, in monopolar pathway, bipolar pathway, and total sample. *p*-values comparing demographic characteristics of patients with monopolar and bipolar stimulation pathways are shown. Polysomnographic data on sleep/wake states is not available for this study. Abbreviations: RNS, responsive neurostimulation; SD, standard deviation.

| | Monopolar pathway (N = 13) | Bipolar pathway (N = 16) | Total sample (N = 25) | *p*-value |
|---|---|---|---|---|
| Age of RNS implantation in years, mean (SD)[a] | 34.4 (11.1) | 43.0 (13.6) | 38.7 (13.9) | 0.15 |
| Female sex, N (%)[b] | 6 (46.2%) | 6 (37.5%) | 12 (48.0%) | 0.72 |
| Seizure focus[b,†] | | | | 0.003* |
| Neocortical, N (%)[b] | 7 (53.8%) | 1 (6.3%) | 8 (32.0%) | 0.009* |
| Mesiotemporal, N (%)[b] | 2 (15.4%) | 11 (68.8%) | 11 (44.0%) | 0.008* |
| Neocortical and mesiotemporal, N (%)[b] | 4 (30.8%) | 4 (25.0%) | 6 (24.0%) | 0.99 |
| Daily electrographic seizure frequency, mean (SD)[a] | 2.6 (5.2) | 6.2 (11.9) | 4.6 (9.6) | 0.51 |
| Epilepsy duration in years, mean (SD)[a] | 22.3 (10.3) | 22.7 (14.5) | 21.6 (13.0) | 0.85 |
| Age of epilepsy onset in years, mean (SD)[a] | 13.8 (7.1) | 21.0 (15.8) | 18.4 (14.0) | 0.25 |
| Etiology of epilepsy, N (%)[b,†] | | | | 0.75 |
| Arteriovenous malformation | 0 (0.0%) | 2 (12.5%) | 1 (8.3%) | NA |
| Congenital nevus syndrome | 0 (0.0%) | 1 (6.3%) | 1 (4.2%) | NA |
| Cryptogenic | 7 (53.8%) | 8 (50.0%) | 13 (54.2%) | NA |
| Encephalitis | 1 (7.7%) | 2 (12.5%) | 3 (12.5%) | NA |
| Perinatal hypoxic ischemic injury | 0 (0.0%) | 1 (6.3%) | 1 (4.2%) | NA |
| Traumatic brain injury | 2 (15.4%) | 2 (12.5%) | 2 (8.3%) | NA |
| Malformation of cortical development | 1 (7.7%) | 0 (0.0%) | 1 (4.2%) | NA |
| Focal cortical dysplasia | 1 (7.7%) | 0 (0.0%) | 1 (4.2%) | NA |
| Periventricular nodular heterotopia | 1 (7.7%) | 0 (0.0%) | 1 (4.2%) | NA |
| 50% responder rate, N (%)[b,†,†] | 9 (69.2%) | 8 (50.0%) | 16 (64.0%) | 0.76[b] |

†Omnibus test;††Responders defined as patients with 50% or greater reduction in seizure frequency, relative to pre-implant baseline; *Significant at 0.05 level. Note: For categorical variables with more than one level, post-hoc pairwise hypothesis testing was conducted only for omnibus tests that were statistically significant at 0.05 level.
   [a] Mann-Whitney *U* test
   [b] Fisher's exact test

*Monopolar stimulation pathway*

State allocations showed convergence to a stationary distribution and minimization of DIC for $K = 2$. For $K > 2$ states, trace plots switched between local optima, suggesting artificial splitting of a single state into multiple states. For interpretation, we thereon refer to the two ordered states as a high and low seizure risk state. Fig. 1 shows an example of the posterior estimates of seizure risk states based on the number of "long episodes" (electrographic seizures) per hour, for a patient with RNS in the monopolar stimulation configuration. The distribution of seizures classified in high and low risk states is shown in Supplementary Figure A2.

Using a monopolar stimulation configuration, the mean stimulation parameter setting over 150,238 person-hours was a charge density of 1.7 ± 0.7 μC/sq cm, pulse frequency of 189.2 ± 54.5 Hz, burst duration of 468.7 ± 1234.7 ms, and pulse width of 155.5 ± 12.7 μs. Stimulation parameters are shown in Table 3 and Fig. 2A. Fig. 3A shows the stimulation parameters associated with transition to a low risk state for the monopolar configuration, along with posterior means and HPD intervals for state-dependent associations. For time points when patients were currently in a low seizure risk state, using a higher charge density, higher stimulation frequency, or shorter burst duration were associated with better chance of remaining in a low seizure risk state (or equivalently, less likely to transition from a low to a high risk state; see Fig. 3A, green circles). For time points when patients were currently in a high seizure risk state, using a lower stimulation frequency (i.e., fewer delivered pulses) or shorter burst duration were associated with a higher probability of transitioning from a high to a low seizure risk state (or equivalently, less likely to remain in a high risk state; see Fig. 3A, red circles, Supplementary Table A1). Robustness of results in the monopolar stimulation configuration was compared to when the subset of four patients with preceding bipolar stimulation were excluded (i.e., with analysis restricted to patients treated only with monopolar stimulation). Subgroup analysis demonstrated consistent associations of higher charge density, higher stimulation frequency, and shorter burst duration with improved chance of remaining in a low seizure risk state with monopolar stimulation, as well as consistent associations between lower stimulation frequency and shorter burst duration with higher probability of transitioning from a high to a low seizure risk state. A near-zero (HPD interval, −0.42 to −0.06) inverse association between charge density and probability of transitioning from a high to low seizure risk state was present when analysis was limited to the subset of patients treated only with monopolar stimulation (Supplementary Figure A3, Supplementary Table A2). We also evaluated robustness to exclusion of patients with malformations of cortical development (MCD), focal cortical dysplasia (FCD), or periventricular nodular heterotopia (PVNH), which may have different tissue electrophysiological properties due to different pathologies. The range of impedances in the three patients with MCD, FCD, and PVNH (mean, 775.6 Ω; range, 565.1–1104.4 Ω) was similar to the rest of the study sample (mean, 671.6 Ω; range, 487.3–1113.0 Ω). Results remained consistent when these patients were excluded. The associations between frequency and burst duration with transition from a high to low risk state remained consistent in direction but became non-significant when patients were excluded, likely due to small sample size (Supplementary Figure A4).

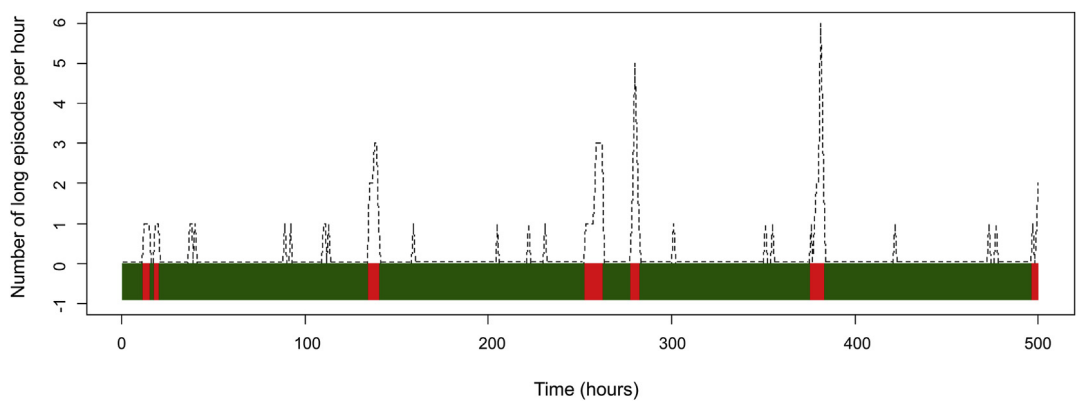

**Fig. 1. Estimated latent seizure risk states for a time sample from an example patient with focal epilepsy implanted with the RNS System.** Number of long episodes per hour is shown in black. Seizure risk state estimates, based on the mode of the posterior distribution of latent states, are shown in red and green. Red bars indicate hours when latent state was identified as a high seizure risk state. Green bars indicate hours when latent state was identified as a low seizure risk state. Abbreviations: RNS, responsive neurostimulation. (For interpretation of the references to colour in this figure legend, the reader is referred to the Web version of this article.)

*Bipolar stimulation pathway*

Using a bipolar stimulation montage, the mean stimulation parameter setting over 130,266 person-hours was a charge density of 2.1 ± 0.8 μC/sq cm, pulse frequency of 198.5 ± 12.1 Hz, burst duration of 100 ms, and pulse width of 157.6 ± 9.5 μs. Fig. 3B shows the stimulation parameters associated with transition to a low risk state for the bipolar stimulation configuration, along with posterior means and HPD intervals. Higher charge densities were associated with better chances of remaining in a low seizure risk state (for times when patients were already in low risk states) as well as better chance of transitioning from a high to low seizure risk state (for times when patients were in high risk states). For times when patients were currently in a low seizure risk state, use of a higher stimulation frequency at 200 Hz was associated with a better chance of remaining in a low risk state, compared to a lower stimulation frequency of 100 Hz (Fig. 3B, green circles). For time points when patients were currently in a high seizure risk state, a lower stimulation frequency of 100 Hz was associated with a better chance of transitioning from a high to low seizure risk state, compared to a higher stimulation frequency of 200 Hz (Fig. 3B, red circles, Supplementary Table A3).

**Table 3**
**RNS stimulation parameters.** The mean, standard deviation (SD), and distribution of parameter values used in monopolar and bipolar stimulation pathways are shown. The range of stimulation parameter values used is also shown, along with the percentage of time points employing each stimulation parameter value. For example, charge density in monopolar pathways ranged from lower values (e.g. 0.5 μC/cm²) to higher values (e.g. 3 μC/cm²), with 9.43% of time points with a charge density of 0.5 μC/cm² and 9.1% of time points with a charge density of 3 μC/cm². Abbreviations: RNS, responsive neurostimulation.

| | Charge density (μC/cm²) | | Frequency (Hz) | | Burst duration (ms) | | Pulse width (μs) | |
|---|---|---|---|---|---|---|---|---|
| | Mean (SD; median) | Stimulation parameter values used (%) | Mean (SD; median) | Stimulation parameter values used (%) | Mean (SD; median) | Stimulation parameter values used (%) | Mean (SD; median) | Stimulation parameter values used (%) |
| Monopolar pathway | 1.7 (0.7; 1.7) | 0.5 (9.43%)<br>0.8 (8.29%)<br>1 (9.45%)<br>1.3 (5.61%)<br>1.4 (2.89%)<br>1.5 (12.84%)<br>1.7 (8.71%)<br>1.8 (3.35%)<br>2 (8.65%)<br>2.3 (12.33%)<br>2.5 (8.97%)<br>2.8 (0.38%)<br>3 (9.1%) | 189.2 (54.5; 200.0) | 2 (7.0%)<br>200 (91.0%)<br>333.3 (2.0%) | 468.7 (1234.7; 100) | 100 (71.0%)<br>200 (16.0%)<br>300 (4.0%)<br>400 (2.0%)<br>5000 (7.0%) | 155.5 (12.7; 160) | 120 (11.0%)<br>160 (89.0%) |
| Bipolar pathway | 2.1 (0.8; 2.0) | 0.5 (1%)<br>0.6 (2.6%)<br>1 (17.4%)<br>1.5 (16.3%)<br>1.9 (1.9%)<br>2 (16.5%)<br>2.3 (1.6%)<br>2.5 (12.8%)<br>3 (30.0%) | 198.5 (12.1; 200.0) | 100 (1.5%)<br>200 (98.5%) | 100 (0; 100) | 100 (100%) | 157.6 (9.5; 160) | 120 (6.0%)<br>160 (94.0%) |

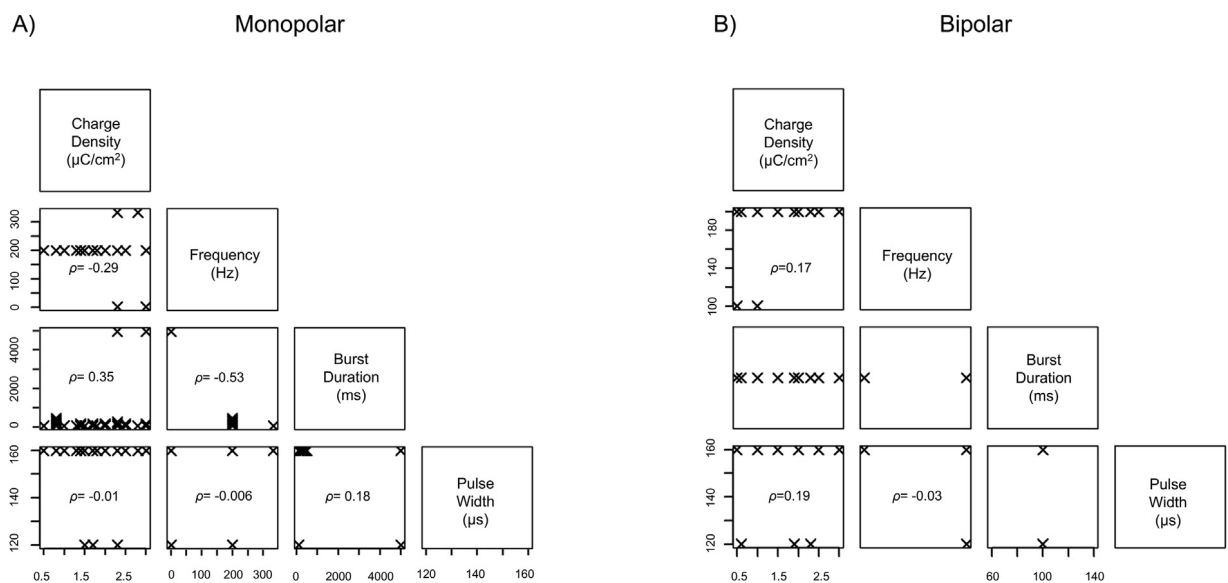

**Fig. 2. RNS stimulation parameters.** Combinations of stimulation parameters used by clinicians within each stimulation pathway are shown. Crosses represent stimulation parameter values. Spearman correlation coefficients ($\rho$) between each pair of stimulation parameters are shown. Abbreviations: RNS, responsive neurostimulation.

*Subgroup analysis: RNS responders and non-responders*

There were 9/13 responders (69.2%) among patients with monopolar stimulation and 8/16 responders (50.0%) among patients with bipolar stimulation (p = 0.76). For both monopolar and bipolar stimulation configurations, RNS responders and non-responders exhibited a similar direction of response to charge density and frequency changes when in low risk states (Supplementary Figures A5-A and A6-A). However, we found that RNS responders and non-responders responded differently to RNS stimulation parameters when in high risk states (Supplementary Figures A5-B and A6-B; yellow squares and black triangles, respectively). For RNS non-responders treated with bipolar stimulation, use of *higher charge densities* in a high risk state was more likely to transition the patient from a high to a low risk state (Supplementary Figure A6-B, black triangles). In comparison, for RNS responders treated with bipolar stimulation, use of *lower stimulation frequencies* was more likely to transition the patient from a high to low risk state (Supplementary Figure A6-B, yellow squares). For monopolar stimulation, non-responders to RNS tended toward opposite responses to increases in charge density and burst duration compared to RNS responders when in high risk states. Whereas RNS responders were more likely to transition from a high to low risk state with low charge density or short burst duration with monopolar stimulation (Supplementary Figure A5-B, yellow squares), RNS non-responders were less likely to transition from a high to low risk state under these adjustments (Supplementary Figure A5-B, black triangles). The association of changes in stimulation frequency with the probability of transitioning the patient from a high to low risk state was significant in the combined patient group but non-significant in subgroup analyses for monopolar stimulation (Supplementary Figure A5-B).

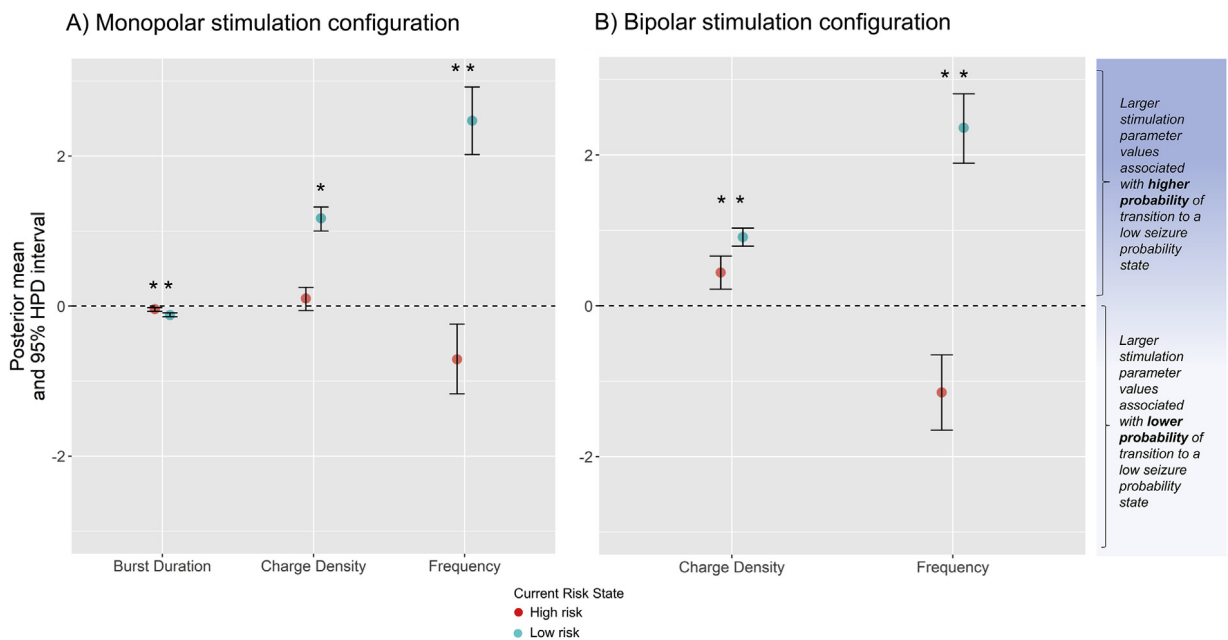

**Fig. 3. Association of RNS stimulation parameters with low future seizure risk, for monopolar and bipolar stimulation configuration.** Posterior means and 95% highest posterior density (HPD) intervals for state-dependent log-odds of association between higher values of each stimulation parameter and probability of transitioning to a low risk state are shown on the y-axis, conditional on current risk state (shown in red and green). (A) Monopolar stimulation: For patients currently in a high seizure risk state (red), shorter burst duration and lower stimulation frequency were associated with transition to a low seizure risk state. For patients currently in a low seizure risk state (green), higher charge density, shorter burst duration, and higher stimulation frequency were associated with remaining in a low seizure risk state. (B) Bipolar stimulation: For patients currently in a high seizure risk state (red), higher charge density and lower stimulation frequency were associated with transition to a low seizure risk state. For patients currently in a low seizure risk state (green), higher charge density and higher stimulation frequency were associated with remaining in a low seizure risk state. Circles correspond to the posterior mean; confidence bands corresponds to HPD intervals. Probabilities are relative to the probability of transitioning to (or remaining in) a high risk state. * = 95% HPD interval does not contain zero. Abbreviations: RNS, responsive neurostimulation. (For interpretation of the references to colour in this figure legend, the reader is referred to the Web version of this article.)

## Discussion

In contemporary practice, RNS stimulation parameters are adjusted empirically and typically held constant for months, a timeframe during which there may be considerable fluctuation in underlying seizure risk. Here, we demonstrate that RNS charge density can modulate seizure risk and that, for the two most common stimulation pathways, the effects of changing other stimulation parameters depend on the current seizure risk state. This suggests that variability in clinical response to RNS may relate at least partly to counterproductive effects of stimulation parameter changes applied during inopportune brain states. These data also suggest that next-generation neurostimulation systems for epilepsy may benefit from real-time feedback and adaptive stimulation capabilities that continuously monitor brain state over prolonged periods of time, which may minimize interpatient variability and enhance clinical outcomes.

Our results provide several key conceptual advances in understanding the relationship between electrical brain stimulation and seizure risk. First, we show that the impact of responsive neurostimulation on modulating seizure frequency may exhibit state-dependence. In particular, we show that stimulation parameters which are effective in one brain state may not be effective, or even have opposite effects, in another brain state. In vitro and in vivo animal models have revealed that the modulatory impact of electrical stimulation is frequency-dependent and timing-dependent [21,30,31], but, to our knowledge, this is the first demonstration in human neuromodulation protocols in epilepsy. A state-based view of electrical stimulation paradigms implies that the impact of neurostimulation extends beyond direct seizure termination, consistent with an emerging concept of RNS having long-term neuromodulatory effects [32] that facilitate transitions from high and low seizure risk states, which may bear an indirect relation to cortical "up" and "down" states. We speculate that these effects may involve state-specific disruption of pathological cortical synchrony.

Second, we discover several new state-dependent effects of various neurostimulation parameters which may catalyze future development of adaptive neurostimulation systems. Currently, the recommended initial settings for the RNS System are: frequency 200 Hz, burst duration 100 ms, charge density 0.5 $\mu C/cm^2$, and pulse width 160 $\mu s$. Over time, current recommendations are to increase charge density stepwise in 0.5 $\mu C/sq$ cm increments [22]. If this fails, burst duration and frequency are often increased, although specific guidelines for these adjustments are not available. Despite these recommendations, empirical evidence supporting these recommendations has not previously been shown. In clinical practice, the most commonly programmed stimulation parameters are an amplitude of 1.5—3.0 mA, pulse width of 160 $\mu s$, burst duration of 100—200 ms, and pulse frequency of 100—200 Hz [33,34]. For both bipolar and monopolar stimulation configurations, we found that higher charge densities were associated with improved outcomes, both for remaining in (bipolar/monopolar) or transitioning to (bipolar) a low seizure risk state. Bipolar stimulation exerts a more focal stimulation effect than monopolar stimulation, such that higher charge densities may make more of a difference for bipolar pathways. In both the real-world outcome (RWO) and pivotal studies [2,35], a similar finding has indirectly been shown. Higher charge densities were often seen with longer duration of treatment in the RWO study, as well as an improvement in median seizure frequency [35]. Whereas duration of treatment may confound correlations between charge density and seizure outcomes in the RWO study, the state-space approach reduces the confounding effect of treatment duration by estimating the association of higher charge densities with the probability of remaining in or transitioning to a lower seizure risk state in the subsequent time point. Although clinicians tended to increase charge density over the timescale of years in the RWO study, decreases in charge density occur either directly or when patients are switched to stimulation pathways involving more cathodes. Interestingly, however, patients in the prior RWO study did better, as a group, than patients in the pivotal trials, although on average charge densities were lower in the RWO study [35]. Whether this is an incidental finding of the RWO study or true association is of interest for future investigation. The relationship between charge density and seizure outcomes is likely more complicated than the early notions that higher charge density necessarily translates into better outcomes, and may relate to factors such as state-dependence and heterogeneity between patient subgroups. Our study has the interesting incidental observation that charge densities applied by clinicians to neocortical leads were lower than the charge densities applied to hippocampal leads. This may not be unexpected; as neocortical stimulation often involves eloquent territories, it is more likely to produce stimulation-triggered signs/symptoms (STS) than mesial temporal stimulation. One protocol that clinicians employ to eliminate or mitigate STS is to reduce charge density, which may result in a lower charge density employed in contemporary practice [36]. It is also possible that lower charge densities may be necessary for response in neocortical regions than in mesiotemporal regions, although larger samples powered for subgroup comparisons are needed to test this hypothesis. Neocortical and mesiotemporal foci have been found to exhibit different responses to stimulation [37], and, presumably, stimulation may also cause different state-dependent effects in different epilepsies. Although comparisons of state-dependent stimulation responses between neocortical and mesiotemporal subgroups are not within scope in this study due to sample size, understanding these differences is of great interest as a future direction of our research.

Within the range of 100—5000 ms, shorter burst durations were associated with remaining in or transitioning to low seizure risk states in monopolar configurations (not evaluated in bipolar configuration due to lack of variability in the bipolar sample). However, depending on whether the patient was currently in a low or high risk state, higher values of stimulation frequency could either lead to transition to a worsening or improvement in seizure risk. Specifically, higher stimulation frequencies were more likely to result in patients remaining in a low seizure risk state, but, if the patient started in a high seizure risk state, a lower stimulation frequency was more likely to transition the patient to a low seizure risk state. This finding was consistent across both bipolar and monopolar stimulation configurations, as well as when the subset of patients with multiple stimulation configurations was excluded.

These results provide evidence to support the empiric recommendation to increase charge density as a first step for patients with subtherapeutic responses to neurostimulation. However, our findings also suggest that the effectiveness of other stimulation parameters, such as stimulation frequency, may depend on the current seizure risk state. Differential state-dependent effects of adjusting frequency may explain why the initial recommended frequency setting for RNS is 200 Hz in RNS but parameter settings usually end up ranging between 100 and 200 Hz. Adaptive approaches to adjusting stimulation parameters may benefit from real-time evaluation of seizure risk state with dynamic online adjustment of stimulation parameters. State-dependent effectiveness of stimulation parameters may suggest a strategy of differentially programming stimulation parameters for different "therapies" (bursts of current pulses) that are delivered within a given detection epoch if redetection of the electrographic seizure occurs. Up to five therapies can be delivered if redetection of the electrographic seizure occurs despite the preceding therapies. Because persistence of a seizure despite repeated therapies is more likely to reflect a high risk state, we postulate that a differential programming strategy may improve effectiveness, with parameters for the first therapy set to those effective in low risk states, and parameters for later therapies set to those effective in high risk states. There are a number of other considerations which must also be taken into account, including whether white/gray matter is stimulated and the location of the stimulation focus [37].

Based on subgroup analysis, we found that, while RNS responders and non-responders tended to respond similarly to parameter changes when in low risk states, non-responders responded differently to stimulation when in a high risk state. Although comparison of RNS responders and non-responders requires validation in matched samples, clinically, this may suggest that some patients may be more likely to respond to charge density and others to stimulation frequency adjustments.

The analytic approach employed in this work has several advantages. First, the latent process approach explicitly models the stochastic nature of seizures, reducing the noise inherent in seizure counting data by probabilistically discriminating between natural variability and sustained changes in risk. Second, the non-homogeneous HMM incorporates associations between RNS stimulation parameters and changes in risk conditional on current state, which allows for inference to be drawn on associations between RNS stimulation parameter changes and fluctuations in seizure risk, holding other RNS stimulation parameters constant. If the association between RNS stimulation parameters and seizure risk is independent of current state, then estimated conditional associations will not significantly differ. This is a major advantage over conventional cross-correlative analyses between stimulation parameter values and seizure risk, which do not account for temporal variation in parameters and do not simultaneously control for other parameters. Third, the temporal directionality relating exogenous inputs to changes in risk allows for causal inference between RNS parameters and changes in risk, with a future risk prediction horizon of 1 hour ahead conditioned on the risk state of the previous hour. Fourth, the Markovian property accounts for memory when identifying seizure risk states. This permits, for example, continuous sequences of zero seizures per hour with only a few interspersed hours with one or two seizures to be identified as low risk despite the presence of a few hours with seizures. Conversely, a continuous sequence of hours with many seizures occurring, with only a few interspersed hours with no seizures, can be identified as high risk despite the potentially random occurrence of a few hours without seizures.

There are several limitations to this study. The patient sample, while spanning a wide range of ages and localizations of epilepsy, was modest in size and may not be representative of all patients implanted with the RNS System. The minimum age evaluated in this sample was 18 years; given increasing interest in RNS use in the pediatric epilepsy population [38,39], generalizability to pediatric populations is of interest. Due to storage limitations, the RNS System does not provide continuous ECoG tracings, so electrographic seizure detections were based on "long episodes," which are an imperfect surrogate for seizures. However, the fact that we only included patients for whom >90% of long-episodes corresponded to electrographic seizures may mitigate this concern. Patients may differ with respect to the number of seizure states that they cycle through. The number of states used was chosen based on the deviance information criterion, which also facilitates interpretability and consistent interpretability in groupwise conclusions. However, if state-dependent electrical stimulation is used in an adaptive stimulation paradigm, it would be beneficial to allow for a varying number of states optimized to the individual patient. Larger samples will be useful for validating these findings across a greater variety of parameter changes, particularly for bipolar or grouped bipolar configurations. A broader range of settings was trialed in the monopolar than bipolar setting in this study. Although we speculate here on translatability to comparing bipolar and monopolar stimulation pathways, small sample sizes and limited matching characteristics caution against direct comparability of these montages and is outside the scope of this study. Larger matched samples are needed for direct comparison of stimulation pathways. Finally, randomized prospective trials evaluating effects against a wider range of randomized stimulation settings or randomized periods of time before stimulation are needed to confirm differences among parameter effects and stimulation configurations. This is an inherent limitation in data collected with neurostimulators programmed to optimize patient outcomes in non-controlled settings. Since bipolar montages tend to be trialed first, differences between bipolar and monopolar montages may reflect differences in timing. In our sample, 4 of the 13 patients with monopolar stimulation had bipolar stimulation first; however, sensitivity analysis demonstrated robustness to exclusion of these patients. Similarly, secondary analyses on differences between neocortical and mesiotemporal subgroups is warranted, as stimulation parameter combinations may not have the same effect on different seizure foci [37]. Recent studies have also found that white and gray matter can have different responses to stimulation, with gray matter stimulation more likely to elicit decreases in high-frequency activity (HFA) power (a proxy for neuronal firing) and white matter stimulation more likely to elicit increases in HFA power [37]. In our study, all electrodes were implanted in gray matter. An interesting extension of our work would be to evaluate whether state-dependent effects also exist in white matter stimulation paradigms (such as those for treatment-resistant depression) [40,41]. This would have implications for adaptive stimulation paradigms for treatment of neuropsychiatric conditions.

In conclusion, our findings provide evidence to support state-dependence in the effectiveness of RNS in modulating seizure frequency. This finding provides a possible explanation for variations in efficacy of RNS electrical stimulation on modulating seizures and suggests the need for incorporating real-time state analysis into adaptive algorithms for future neurostimulation systems.

## Funding

Publication made possible through support from the University of California, San Francisco (UCSF) Division of Epilepsy and UCSF Open Access Publishing Fund. ANK is supported by the Citizens United for Research In Epilepsy (CURE): Taking Flight Award. ETW is supported by a fellowship from the Gulf Coast Consortia, on the NLM Training Program in Biomedical Informatics and Data Science T15LM007093. VRR is supported by the Ernest Gallo Distinguished Professorship at UCSF.

## CRediT authorship contribution statement

**Sharon Chiang:** Conceptualization, Methodology, Software, Formal analysis, Investigation, Data interpretation, Writing - original draft. **Ankit N. Khambhati:** Conceptualization, Data curation, Investigation, Writing - review & editing. **Emily T. Wang:** Software, Writing - review & editing. **Marina Vannucci:** Software, Writing - review & editing. **Edward F. Chang:** Data interpretation, Writing - review & editing. **Vikram R. Rao:** Conceptualization, Investigation, Data interpretation, Resources, Data curation, Writing - review & editing, Supervision.

## Declaration of competing interest

VRR has served as a consultant for NeuroPace, Inc., manufacturer of the RNS System, but declares no targeted compensation or other support for this study. All other authors declare no conflicts of interest relevant to this study.

## Acknowledgements

The authors thank SeizureTracker.com for collaboration on projects that contributed to code development.

## Appendix A. Supplementary data

Supplementary data to this article can be found online at https://doi.org/10.1016/j.brs.2021.01.023.

## References

[1] Bergey GK, Morrell MJ, Mizrahi EM, Goldman A, King-Stephens D, Nair D, et al. Long-term treatment with responsive brain stimulation in adults with refractory partial seizures. Neurology 2015;84(8):810–7.

[2] Nair DR, Laxer KD, Weber PB, Murro AM, Park YD, Barkley GL, et al. Nine-year prospective efficacy and safety of brain-responsive neurostimulation for focal epilepsy. Neurology 2020;95(9):e1244–56.

[3] Geller EB, Skarpaas TL, Gross RE, Goodman RR, Barkley GL, Bazil CW, et al. Brain-responsive neurostimulation in patients with medically intractable mesial temporal lobe epilepsy. Epilepsia 2017;58(6):994–1004.

[4] Ma BB, Rao VR. Responsive neurostimulation: candidates and considerations. Epilepsy Behav 2018;88:388–95.

[5] Ma BB, Fields MC, Knowlton RC, Chang EF, Szaflarski JP, Marcuse LV, et al. Responsive neurostimulation for regional neocortical epilepsy. Epilepsia 2020;61(1):96–106.

[6] Kossoff EH, Ritzl EK, Politsky JM, Murro AM, Smith JR, Duckrow RB, et al. Effect of an external responsive neurostimulator on seizures and electrographic discharges during subdural electrode monitoring. Epilepsia 2004;45(12):1560–7.

[7] Lesser RP, Kim SH, Beyderman L, Miglioretti DL, Webber WR, Bare M, et al. Brief bursts of pulse stimulation terminate afterdischarges caused by cortical stimulation. Neurology 1999;53(9):2073–81.

[8] Skarpaas TL, Morrell MJ. Intracranial stimulation therapy for epilepsy. Neurotherapeutics 2009;6(2):238–43.

[9] Chkhenkeli SA, Chkhenkeli IS. Effects of therapeutic stimulation of nucleus caudatus on epileptic electrical activity of brain in patients with intractable epilepsy. Stereotact Funct Neurosurg 1997;69(1–4 Pt 2):221–4.

[10] Cordeiro JG, Somerlik KH, Cordeiro KK, Aertsen A, Araujo JC, Schulze-Bonhage A. Modulation of excitability by continuous low- and high-frequency stimulation in fully hippocampal kindled rats. Epilepsy Res 2013;107(3):224–30.

[11] Chiang S, Vannucci M, Goldenholz DM, Moss R, Stern JM. Epilepsy as a dynamic disease: a Bayesian model for differentiating seizure risk from natural variability. Epilepsia Open 2018;3(2):236–46.

[12] Chiang S, Goldenholz DM, Moss R, Rao VR, Haneef Z, Theodore WH, et al. Prospective validation study of an epilepsy seizure risk system for outpatient evaluation. Epilepsia 2019;61(1):29–38.

[13] Baud MO, Proix T, Rao VR, Schindler K. Chance and risk in epilepsy. Curr Opin Neurol 2020;33(2):163–72.

[14] Proix T, Truccolo W, Leguia MG, Tcheng TK, King-Stephens D, Rao VR, Baud MO. Forecasting seizure risk in adults with focal epilepsy: a development and validation study. Lancet Neurol. 2021 Feb;20(2):127–35. https://doi.org/10.1016/S1474-4422(20)30396-3.

[15] Karoly PJ, Cook MJ, Maturana M, Nurse ES, Payne D, Brinkmann BH, et al. Forecasting cycles of seizure likelihood. Epilepsia 2020;61(4):776–86.

[16] Karoly PJ, Goldenholz DM, Freestone DR, Moss RE, Grayden DB, Theodore WH, et al. Circadian and circaseptan rhythms in human epilepsy: a retrospective cohort study. Lancet Neurol 2018;17(11):977–85.

[17] Karoly PJ, Ung H, Grayden DB, Kuhlmann L, Leyde K, Cook MJ, et al. The circadian profile of epilepsy improves seizure forecasting. Brain 2017;140(8):2169–82.

[18] Anderson CT, Tcheng TK, Sun FT, Morrell MJ. Day-night patterns of epileptiform activity in 65 patients with long-term ambulatory electrocorticography. J Clin Neurophysiol 2015;32(5):406–12.

[19] Baud MO, Kleen JK, Mirro EA, Andrechak JC, King-Stephens D, Chang EF, et al. Multi-day rhythms modulate seizure risk in epilepsy. Nat Commun 2018;9(1):88.

[20] Wester JC, Contreras D. Differential modulation of spontaneous and evoked thalamocortical network activity by acetylcholine level in vitro. J Neurosci 2013;33(45):17951–66.

[21] Li G, Henriquez CS, Frohlich F. Unified thalamic model generates multiple distinct oscillations with state-dependent entrainment by stimulation. PLoS Comput Biol 2017;13(10):e1005797.

[22] NeuroPace RNS. System User Manual. CA: NeuroPace Mountain View; 2015.

[23] Spencer DC, Sun FT, Brown SN, Jobst BC, Fountain NB, Wong VS, et al. Circadian and ultradian patterns of epileptiform discharges differ by seizure-onset location during long-term ambulatory intracranial monitoring. Epilepsia 2016;57(9):1495–502.

[24] Leguia MG, Andrzejak RG, Rummel C, et al. Seizure cycles in focal epilepsy. JAMA Neurol 2021. https://doi.org/10.1001/jamaneurol.2020.5370. Published online February 08.

[25] Gelman A, Carlin JB, Stern HS, Dunson DB, Vehtari A, Rubin DB. Bayesian data analysis. 2nd ed. Chapman & Hall/CRC; 2013.

[26] Spiegelhalter DJ, Best NG, Carlin BP, Van Der Linde A. Bayesian measures of model complexity and fit. J Roy Stat Soc B 2002;64(4):583–639.

[27] de Bragança Pereira CA, Stern JM. Evidence and credibility: full Bayesian significance test for precise hypotheses. J Entropy 1999;1(4):99–110.

[28] Heck CN, King-Stephens D, Massey AD, Nair DR, Jobst BC, Barkley GL, et al. Two-year seizure reduction in adults with medically intractable partial onset epilepsy treated with responsive neurostimulation: final results of the RNS System Pivotal trial. Epilepsia 2014;55(3):432–41.

[29] Krucoff MO, Wozny TA, Lee AT, Rao VR, Chang EF. Operative technique and lessons learned from surgical implantation of the NeuroPace responsive Neurostimulation® system in 57 consecutive patients. Oper Neurosurg (Hagerstown) 2021;20(2):E98–109.

[30] Sadowski JH, Jones MW, Mellor JR. Sharp-wave ripples orchestrate the induction of synaptic plasticity during reactivation of place cell firing patterns in the hippocampus. Cell Rep 2016;14(8):1916–29.

[31] Tremblay SA, Chapman CA, Courtemanche R. State-dependent entrainment of prefrontal cortex local field potential activity following patterned stimulation of the cerebellar vermis. Front Syst Neurosci 2019;13:60.

[32] Kokkinos V, Sisterson ND, Wozny TA, Richardson RM. Association of closed-loop brain stimulation neurophysiological features with seizure control among patients with focal epilepsy. JAMA Neurol 2019;76(7):800–8.

[33] Geller EB. Responsive neurostimulation: review of clinical trials and insights into focal epilepsy. Epilepsy Behav 2018;88:11–20.

[34] Sun FT, Morrell MJ. Closed-loop neurostimulation: the clinical experience. Neurotherapeutics 2014;11(3):553–63.

[35] Razavi B, Rao VR, Lin C, Bujarski KA, Patra SE, Burdette DE, et al. Real-world experience with direct brain-responsive neurostimulation for focal onset seizures. Epilepsia 2020;61(8):1749–57.

[36] Hixon AM, Brown MG, McDermott D, Destefano S, Abosch A, Kahn L, et al. RNS modifications to eliminate stimulation-triggered signs or symptoms (STS): case series and practical guide. Epilepsy Behav 2020;112:107327.

[37] Mohan UR, Watrous AJ, Miller JF, Lega BC, Sperling MR, Worrell GA, et al. The effects of direct brain stimulation in humans depend on frequency, amplitude, and white-matter proximity. Brain Stimulation 2020;13(5):1183–95.

[38] Singhal NS, Numis AL, Lee MB, Chang EF, Sullivan JE, Auguste KI, et al. Responsive neurostimulation for treatment of pediatric drug-resistant epilepsy. Epilepsy Behav Case Rep 2018;10:21–4.

[39] Kokoszka MA, Panov F, La Vega-Talbott M, McGoldrick PE, Wolf SM, Ghatan S. Treatment of medically refractory seizures with responsive neurostimulation: 2 pediatric cases. J Neurosurg Pediatr 2018;21(4):421–7.

[40] Riva-Posse P, Choi KS, Holtzheimer PE, McIntyre CC, Gross RE, Chaturvedi A, et al. Defining critical white matter pathways mediating successful subcallosal cingulate deep brain stimulation for treatment-resistant depression. Biol Psychiatr 2014;76(12):963–9.

[41] Davidson B, Lipsman N, Meng Y, Rabin JS, Giacobbe P, Hamani C. The use of tractography-based targeting in deep brain stimulation for psychiatric indications. Front Hum Neurosci 2020;14:588423.